\documentclass[twocolumn]{aastex61}

\newcommand\aastex{AAS\TeX}

\received{2017 October 13}
\revised{2018 January 18}
\accepted{2018 January 22}
\submitjournal{ApJL}

\shorttitle{\aastex\ Type-Ic SN~2017dio interacting with CSM}
\shortauthors{Kuncarayakti et al.}
\begin{document}

\title{SN~2017dio: a type-Ic supernova exploding in a hydrogen-rich circumstellar medium\footnote{Based on observations made with the NOT, operated by the Nordic Optical Telescope Scientific Association at the Observatorio del Roque de los Muchachos, La Palma, Spain, of the Instituto de Astrofisica de Canarias.
This work is based (in part) on observations collected at the European Organisation for Astronomical Research in the Southern Hemisphere, Chile as part of PESSTO, (the Public ESO Spectroscopic Survey for Transient Objects Survey) ESO program 188.D-3003, 191.D-0935, 197.D-1075.
Based on observations made with the Liverpool Telescope operated on the island of La Palma by Liverpool John Moores University in the Spanish Observatorio del Roque de los Muchachos of the Instituto de Astrofisica de Canarias with financial support from the UK Science and Technology Facilities Council. }}

\correspondingauthor{Hanindyo Kuncarayakti}
\email{hanindyo.kuncarayakti@utu.fi}

%\author[0000-0002-0786-7307]{Greg J. Schwarz}
%\affil{American Astronomical Society \\
%2000 Florida Ave., NW, Suite 300 \\
%Washington, DC 20009-1231, USA}

\author[0000-0002-1132-1366]{Hanindyo Kuncarayakti}
\affiliation{Finnish Centre for Astronomy with ESO (FINCA), University of Turku, V\"{a}is\"{a}l\"{a}ntie 20, 21500 Piikki\"{o}, Finland}
\affiliation{Tuorla Observatory, Department of Physics and Astronomy, University of Turku, V\"{a}is\"{a}l\"{a}ntie 20, 21500 Piikki\"{o}, Finland}
%\collaboration{(AAS Journals Data Scientists collaboration)}

\author{Keiichi Maeda}
\affiliation{Department of Astronomy, Graduate School of Science, Kyoto University, Sakyo-ku, Kyoto 606-8502, Japan}
\affiliation{Kavli Institute for the Physics and Mathematics of the Universe (WPI), The University of Tokyo, 5-1-5 Kashiwanoha, Kashiwa, Chiba 277-8583, Japan}

\author{Christopher J. Ashall}
\affiliation{Astrophysics Research Institute, Liverpool John Moores University, IC2, Liverpool Science Park, 146 Brownlow Hill, Liverpool L3 5RF, UK}
\affiliation{Department of Physics, Florida State University, Tallahassee, FL 32306, USA}

\author{Simon J. Prentice}
\affiliation{Astrophysics Research Institute, Liverpool John Moores University, IC2, Liverpool Science Park, 146 Brownlow Hill, Liverpool L3 5RF, UK}

\author{Seppo Mattila}
%\affiliation{Finnish Centre for Astronomy with ESO (FINCA), University of Turku, V\"{a}is\"{a}l\"{a}ntie 20, 21500 Piikki\"{o}, Finland}
\affiliation{Tuorla Observatory, Department of Physics and Astronomy, University of Turku, V\"{a}is\"{a}l\"{a}ntie 20, 21500 Piikki\"{o}, Finland}

\author{Erkki Kankare}
\affiliation{Astrophysics Research Centre, School of Mathematics and Physics, Queen's University Belfast, Belfast BT7 1NN, UK}

\author{Claes Fransson}
\affiliation{Department of Astronomy and The Oskar Klein Centre, AlbaNova University Center, Stockholm University, SE-10691 Stockholm, Sweden}

\author{Peter Lundqvist}
\affiliation{Department of Astronomy and The Oskar Klein Centre, AlbaNova University Center, Stockholm University, SE-10691 Stockholm, Sweden}

\author{Andrea Pastorello}
\affiliation{INAF-Osservatorio Astronomico di Padova, Vicolo dell'Osservatorio 5, I-35122 Padova, Italy}

\author{Giorgos Leloudas}
\affiliation{Dark Cosmology Centre, Niels Bohr Institute, University of Copenhagen, Juliane Maries vej 30, 2100 Copenhagen, Denmark}

\author{Joseph P. Anderson}
\affiliation{European Southern Observatory, Alonso de C\'ordova 3107, Vitacura, Casilla 19001, Santiago, Chile}

\author{Stefano Benetti}
\affiliation{INAF-Osservatorio Astronomico di Padova, Vicolo dell’Osservatorio 5, I-35122 Padova, Italy}

\author{Melina C. Bersten}
\affiliation{Facultad de Ciencias Astron\'omicas y Geof\'isicas, Universidad Nacional de La Plata, Paseo del Bosque S/N, B1900FWA La Plata,
Argentina}
\affiliation{Instituto de Astrof\'isica de La Plata (IALP), CONICET, Argentina}
\affiliation{Kavli Institute for the Physics and Mathematics of the Universe (WPI), The University of Tokyo, 5-1-5 Kashiwanoha, Kashiwa, Chiba 277-8583, Japan}

\author{Enrico Cappellaro}
\affiliation{INAF-Osservatorio Astronomico di Padova, Vicolo dell’Osservatorio 5, I-35122 Padova, Italy}

\author{R\'egis Cartier}
\affiliation{Cerro Tololo Inter-American Observatory, National Optical Astronomy Observatory, Casilla 603, La Serena, Chile}

\author{Larry Denneau}
\affiliation{Institute for Astronomy, University of Hawaii, 2680 Woodlawn Dr., Honolulu, HI 96822}

\author{Massimo Della Valle}
\affiliation{INAF-Napoli, Osservatorio Astronomico di Capodimonte, Salita Moiariello 16, I-80131 Napoli, Italy}
\affiliation{International Center for Relativistic Astrophysics, Piazzale della Repubblica 2, I-65122, Pescara, Italy}

\author{Nancy Elias-Rosa}
\affiliation{INAF-Osservatorio Astronomico di Padova, Vicolo dell’Osservatorio 5, I-35122 Padova, Italy}

\author{Gast\'on Folatelli}
\affiliation{Facultad de Ciencias Astron\'omicas y Geof\'isicas, Universidad Nacional de La Plata, Paseo del Bosque S/N, B1900FWA La Plata,
Argentina}
\affiliation{Instituto de Astrof\'isica de La Plata (IALP), CONICET, Argentina}
\affiliation{Kavli Institute for the Physics and Mathematics of the Universe (WPI), The University of Tokyo, 5-1-5 Kashiwanoha, Kashiwa, Chiba 277-8583, Japan}

\author{Morgan Fraser}
\affiliation{School of Physics, O’Brien Centre for Science North, University College Dublin, Belfield, Dublin 4, Ireland}
\affiliation{Royal Society - Science Foundation Ireland University Research Fellow}

\author{Llu\'{i}s Galbany}
\affiliation{PITT PACC, Department of Physics and Astronomy, University of Pittsburgh, Pittsburgh, PA 15260, USA}

\author{Christa Gall}
\affiliation{Department of Physics and Astronomy, Aarhus University, Ny Munkegade 120, DK-8000 Aarhus C, Denmark}
\affiliation{Dark Cosmology Centre, Niels Bohr Institute, University of Copenhagen, Juliane Maries vej 30, 2100 Copenhagen, Denmark}

\author{Avishay Gal-Yam}
\affiliation{Benoziyo Center for Astrophysics, Weizmann Institute of Science, 76100 Rehovot, Israel}

\author{Claudia P. Guti\'errez}
\affiliation{Department of Physics and Astronomy, University of Southampton, Southampton, SO17 1BJ, UK}

\author{Aleksandra Hamanowicz}
\affiliation{European Southern Observatory, Karl-Schwarzschild Str. 2, 85748 Garching bei Munchen, Germany}
\affiliation{Warsaw University Astronomical Observatory, Al. Ujazdowskie 4, 00-478 Warszawa, Poland}

\author{Ari Heinze}
\affiliation{Institute for Astronomy, University of Hawaii, 2680 Woodlawn Dr., Honolulu, HI 96822}

\author{Cosimo Inserra}
\affiliation{Astrophysics Research Centre, School of Mathematics and Physics, Queen's University Belfast, Belfast BT7 1NN, UK}
\affiliation{Department of Physics \& Astronomy, University of Southampton, Southampton, Hampshire, SO17 1BJ, UK}

\author{Tuomas Kangas}
\affiliation{Space Telescope Science Institute, 3700 San Martin Drive, Baltimore, MD 21218, USA}

\author{Paolo Mazzali}
\affiliation{Astrophysics Research Institute, Liverpool John Moores University, IC2, Liverpool Science Park, 146 Brownlow Hill, Liverpool L3 5RF, UK}
\affiliation{Max-Planck-Institut f\"ur Astrophysik, Karl-Schwarzschild-Str. 1, D-85748 Garching, Germany}

\author{Andrea Melandri}
\affiliation{INAF - Osservatorio Astronomico di Brera, via E. Bianchi 46, I-23807, Merate (LC), Italy}

\author{Giuliano Pignata}
\affiliation{Departamento de Ciencias Fisicas, Universidad Andres Bello, Avda. Republica 252, Sazi\'e, 2320, Santiago, Chile}
\affiliation{Nuncio Monse\~{n}or S\'otero Sanz 100, Providencia, Santiago, Chile}

\author{Armin Rest}
\affiliation{Space Telescope Science Institute, 3700 San Martin Drive, Baltimore, MD 21218, USA}
\affiliation{ Department of Physics and Astronomy, The Johns Hopkins University, 3400 North Charles Street, Baltimore, MD 21218, USA}

\author{Thomas Reynolds}
\affiliation{Tuorla Observatory, Department of Physics and Astronomy, University of Turku, V\"{a}is\"{a}l\"{a}ntie 20, 21500 Piikki\"{o}, Finland}

\author{Rupak Roy}
\affiliation{Department of Astronomy and The Oskar Klein Centre, AlbaNova University Center, Stockholm University, SE-10691 Stockholm, Sweden}

\author{Stephen J. Smartt}
\affiliation{Astrophysics Research Centre, School of Mathematics and Physics, Queen's University Belfast, Belfast BT7 1NN, UK}

\author{Ken W. Smith}
\affiliation{Astrophysics Research Centre, School of Mathematics and Physics, Queen's University Belfast, Belfast BT7 1NN, UK}

\author{Jesper Sollerman}
\affiliation{Department of Astronomy and The Oskar Klein Centre, AlbaNova University Center, Stockholm University, SE-10691 Stockholm, Sweden}

\author{Auni Somero}
\affiliation{Institut de Física d'Altes Energies (IFAE). Edifici Cn, Universitat Aut\`onoma de Barcelona (UAB), E-08193 Bellaterra (Barcelona), Spain}
\affiliation{Tuorla Observatory, Department of Physics and Astronomy, University of Turku, V\"{a}is\"{a}l\"{a}ntie 20, 21500 Piikki\"{o}, Finland}

\author{Brian Stalder}
\affiliation{LSST, 950 N. Cherry Ave, Tucson, AZ 85719, USA}

\author{Maximilian Stritzinger}
\affiliation{Department of Physics and Astronomy, Aarhus University, Ny Munkegade 120, DK-8000 Aarhus C, Denmark}

\author{Francesco Taddia}
\affiliation{Department of Astronomy and The Oskar Klein Centre, AlbaNova University Center, Stockholm University, SE-10691 Stockholm, Sweden}

\author{Lina Tomasella}
\affiliation{INAF-Osservatorio Astronomico di Padova, Vicolo dell’Osservatorio 5, I-35122 Padova, Italy}

\author{John Tonry}
\affiliation{Institute for Astronomy, University of Hawaii, 2680 Woodlawn Dr., Honolulu, HI 96822}

\author{Henry Weiland}
\affiliation{Institute for Astronomy, University of Hawaii, 2680 Woodlawn Dr., Honolulu, HI 96822}

\author{David R. Young}
\affiliation{Astrophysics Research Centre, School of Mathematics and Physics, Queen's University Belfast, Belfast BT7 1NN, UK}

%\author{Avishay Gal-Yam}
%Elias-Rosa, Gall, Reynolds, Capellaro

%\author{NUTS}
%\author{ePESSTO}

%\nocollaboration

%\altaffiliation{Creator of AASTeX v6.1}
%\collaboration{(LaTeX collaboration)}
%\affiliation{AAS Director of Publishing}
%\affiliation{American Astronomical Society \\
%2000 Florida Ave., NW, Suite 300 \\
%Washington, DC 20009-1231, USA}
%
%\author{Jeff Lewandowski}
%\affiliation{IOP Senior Publisher for the AAS Journals}
%\affiliation{IOP Publishing, Washington, DC 20005}



\begin{abstract}

SN~2017dio shows both spectral characteristics of a type-Ic supernova (SN) and signs of a hydrogen-rich circumstellar medium (CSM). 
Prominent, narrow emission lines of H and He are superposed on the continuum. 
Subsequent evolution revealed that the SN ejecta are interacting with the CSM. 
The initial SN Ic identification was confirmed by removing the CSM interaction component from the spectrum and comparing with known SNe Ic, and reversely, adding a CSM interaction component to the spectra of known SNe Ic and comparing them to SN~2017dio. Excellent agreement was obtained with both procedures, reinforcing the SN Ic classification. 
{The light curve constrains the pre-interaction SN Ic peak absolute magnitude to be around $M_g = -17.6$ mag. No evidence of significant extinction is found, ruling out a brighter luminosity required by a SN Ia classification.
}
These pieces of evidence support the view that SN~2017dio is a SN Ic, and therefore the first firm case of a SN Ic with signatures of hydrogen-rich CSM in the early {spectrum}.
The CSM is unlikely to have been shaped by steady-state stellar winds.
{The mass loss of the progenitor star must have been intense, 
$\dot{M} \sim 0.02$ $(\epsilon_{H\alpha}/0.01)^{-1}$ $(v_\textrm{wind}/500$ km s$^{-1}$) $(v_\textrm{shock}/10 000$ km s$^{-1})^{-3}$ $M_\odot$~yr$^{-1}$}, peaking at a few decades before the SN. Such {a} high mass loss {rate} might have been experienced by the progenitor through eruptions or binary stripping.

%
%This example manuscript is intended to serve as a tutorial and template for
%authors to use when writing their own AAS Journal articles. The manuscript
%includes a history of \aastex\ and documents the new features in the
%previous version, 6.0, as well as the new features in version 6.1. This
%manuscript includes many figure and table examples to illustrate these new
%features.  Information on features not explicitly mentioned in the article
%can be viewed in the manuscript comments or more extensive online
%documentation. Authors are welcome replace the text, tables, figures, and
%bibliography with their own and submit the resulting manuscript to the AAS
%Journals peer review system.  The first lesson in the tutorial is to remind
%authors that the AAS Journals, the Astrophysical Journal (ApJ), the
%Astrophysical Journal Letters (ApJL), and Astronomical Journal (AJ), all
%have a 250 word limit for the abstract.  If you exceed this length the
%Editorial office will ask you to shorten it.

\end{abstract}




\section{Introduction} 

%At the end of the evolution, massive stars explode as core-collapse supernovae (SNe). 
Core-collapse supernovae (SNe) mark the endpoints of {the evolution of massive stars}.
Due to mass loss, either via winds \citep{vink01} or close binary interaction \citep{podsiadlowski92}, some of these stars end up with only a small amount of envelope left. They are thought to be the progenitors of stripped-envelope (SE) SNe {including} type-Ic, Ib, and IIb SNe {\citep[][]{galyam16}}. SNe~Ic are considered to be the most highly stripped as they lack both H and He lines in their spectrum \citep[{e.g.}][]{prentice17}. 
{Therefore, their} progenitors must have experienced significant mass loss before the time of the explosion. 

SNe IIn \citep{schlegel90} and Ibn \citep{pastorello07} show narrow emission lines of H and He, respectively, and are thought to be interacting with a dense circumstellar medium (CSM) created through intensive mass loss of the progenitor. 
{On the other hand, most SNe Ib/c do not show clear signs of CSM}. 
Some {examples} of {SNe Ib/c with signs of CSM} include SNe 2014C \citep{milisav15}, 2010mb \citep{benami14}, and 2001em \citep{chugai06}. 
Late-time H$\alpha$ emission has been detected in several {luminous \citep{roy16} and superluminous SNe Ic \citep{yan17}}, and in a few SNe Ib/Ic \citep{vinko17}.
However, H emission lines have \emph{never} been observed in the early phases of SNe Ic.

SN~2017dio was discovered on 2017-04-26\footnote{Dates are UTC throughout the paper.} by ATLAS\footnote{The Asteroid Terrestrial-impact Last Alert System \citep{tonry11}.} at 18.29 mag (cyan-band). The last non-detection by ATLAS was on 2017-03-29, $>19.76$ mag (cyan).
Subsequent spectroscopy by ePESSTO\footnote{Extended Public ESO Spectroscopic Survey of Transient Objects \citep{smartt15}.} on 2017-04-29 suggests that the spectrum of SN~2017dio resembles those of SNe Ic \citep{cartier17}. The classification spectrum also shows narrow Balmer emission lines superposed on the SN Ic spectrum. 
These lines yield a redshift of $z = 0.037$ that corresponds to a {distance modulus of 36.0 mag, assuming $H_0 = 72$~km~s$^{-1}$~Mpc$^{-1}$}. 
Line-of-sight Milky Way extinction is $E(B-V) = 0.028$~mag \citep{schlafly11}.
Following discovery and classification, we triggered further photometric and spectroscopic observations within the ePESSTO and NUTS\footnote{Nordic Optical Telescope Unbiased Transient Survey \citep{mattila16}.} collaborations.
%Here we report the results of our observations, and discuss the physical interpretation of the data.

\section{Observations and data reduction}
\subsection{Photometry}
Optical photometry was obtained in \textit{ugriz} filters using ALFOSC\footnote{Andalucia Faint Object Spectrograph and Camera.} at the Nordic Optical Telescope (NOT), IO:O at the Liverpool Telescope (LT), and Spectral at the 2-m telescopes of Las Cumbres Observatory. 
{\textit{JHK} photometry was obtained using NOT/NOTCam at one epoch, 2017-07-27.}

Standard reductions techniques were applied using \textsc{Iraf}\footnote{\href{http://iraf.net/}{http://iraf.net/}}, and aperture photometry was carried out for the SN and surrounding stars. Photometric zero points were derived by comparing the instrumental magnitudes of the stars in the field to  photometry from Sloan Digital Sky Survey, Data Release 14\footnote{\href{http://www.sdss.org/dr14/}{http://www.sdss.org/dr14/}}.
Additionally, pre-discovery \textit{V}-band data points were obtained from the CRTS\footnote{Catalina Real-Time Transient Survey \citep{drake09}.} archive and these were scaled to have the data point at 2017-05-01 match the \textit{g}-band data point at the same epoch.

\subsection{Spectroscopy}
Optical slit-spectroscopy was obtained using ALFOSC at the NOT,  EFOSC2\footnote{ESO Faint Object Spectrograph and Camera 2 \citep{buzzoni84}} at the ESO New Technology Telescope (NTT, through the ePESSTO programme), and SPRAT\footnote{Spectrograph for the Rapid Acquisition of Transients \citep{piascik14}} at the LT.
ALFOSC spectra cover 3300-9500~\AA, with a resolution of 16~\AA. With EFOSC2, {grism} \#13, the coverage was 3500-9300~\AA, at a 21~\AA~resolution. Additionally, in two epochs the EFOSC2 spectra were taken using {grism} \#11 (16~\AA~resolution). SPRAT covers 4000-8000~\AA, with a 18~\AA~resolution.

After the standard reduction procedures, the spectra were extracted, then wavelength and flux calibrated. \textsc{Iraf}, \textsc{Foscgui}\footnote{\href{http://sngroup.oapd.inaf.it/foscgui.html}{http://sngroup.oapd.inaf.it/foscgui.html}}, and PESSTO pipeline\footnote{\href{https://github.com/svalenti/pessto}{https://github.com/svalenti/pessto}} tools were used in the procedures.

\section{Results and discussion}
\subsection{Spectra and evolution}

\begin{figure*}
\centering
%\plotone{vwspec.eps}
%\includegraphics[width=.8\linewidth]{f1_vwspec2.eps}
\includegraphics[width=.9\linewidth]{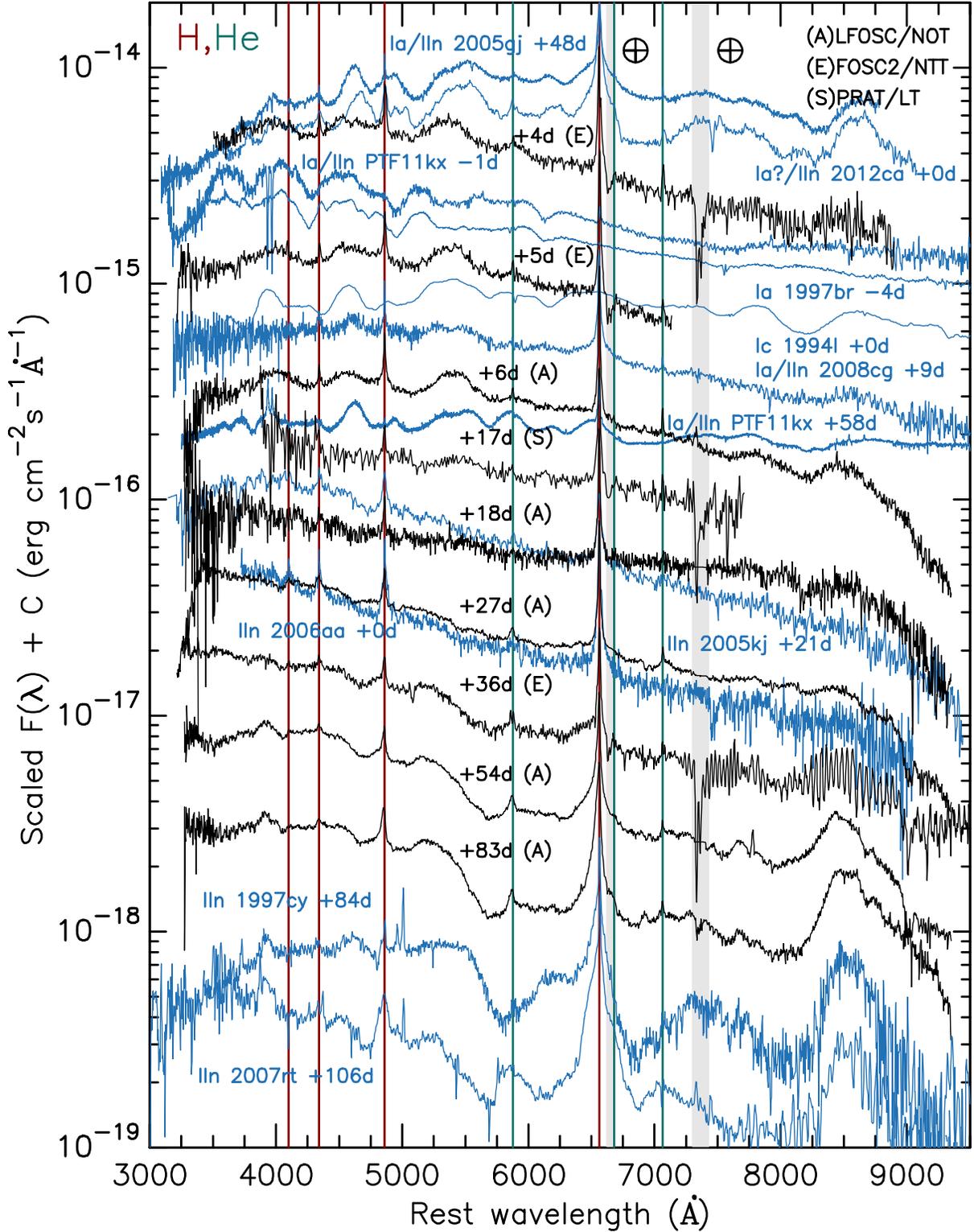}
\caption{
Spectral sequence of SN~2017dio. 
{Comparison spectra of other SNe (obtained through WISeREP\footnote{\href{https://wiserep.weizmann.ac.il/}{https://wiserep.weizmann.ac.il/}}, \citealt{yaron12}) are plotted in blue. 
Phases for SN~2017dio are days after the discovery, while for other SNe phases are post maximum. }
The rest wavelengths of the H~I lines are indicated with dark red and He~I with dark green lines. Telluric absorption regions in SN~2017dio spectra are indicated with grey shade.
{SN~2017dio spectra will be made publicly available at WISeREP {and the Open Supernova Catalog\footnote{\href{https://sne.space/}{https://sne.space/}} \citep{guillochon17}}.}
}
\label{specs}
\end{figure*}

At four days post-discovery, SN~2017dio was classified as a SN Ic. The subsequent spectra at +5 and +6~days (+0 day being the discovery date), show very similar appearance to the classification spectrum (Figure~\ref{specs}). The spectra are relatively smooth, with broad features {at wavelength ranges} $\sim$4000-5500 and $\sim$8000-9000~\AA. There are emission lines of H, and He I at $\lambda\lambda$5876, 7065. At +18~days, the continuum becomes almost featureless and after this point the spectrum starts to show evolving features. Both H and He emission lines are always present, and are resolved by our instruments, with Lorentzian full-width at half-maximum (FWHM) corresponding to an expansion velocity of $\sim500$ km~s$^{-1}$ (corrected for instrumental broadening).
These lines are therefore not originating from an underlying H~II region and are more likely attributed to the SN CSM.

\begin{figure*}
\centering
\includegraphics[width=\linewidth]{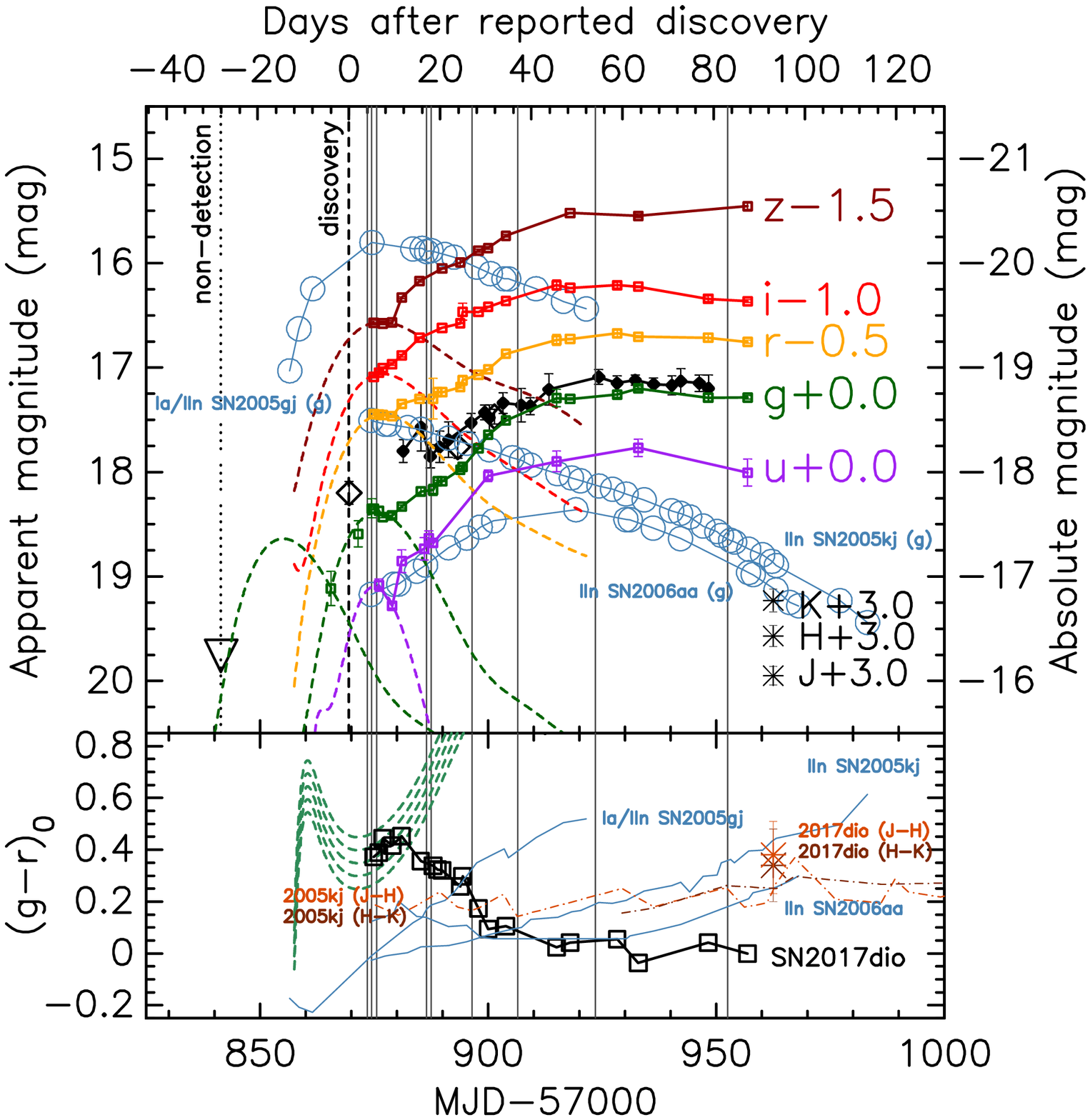}
\caption{
{\textit{(Top)}}
Light curves of SN~2017dio in \textit{ugriz} bands, {and \textit{JHK} photometric points}. 
CRTS data points are {plotted as \textit{g}-band but} not connected to the \textit{g}-band LC.
The absolute magnitudes assume only the distance modulus of {36.0} mag.
%The epochs of ATLAS non-detection and discovery are indicated with dotted and dashed vertical lines, respectively. 
ATLAS cyan-band ({open black diamonds}) and orange-band ({filled} black diamonds) data points are plotted. {A} non-detection ($3\sigma$) limiting magnitude is plotted with upside-down triangle. 
Vertical grey lines indicate the epochs of spectroscopy. 
Dashed curves represent template SNe Ic LCs of \citet{taddia15}. 
%The same X-axis shift was applied to the \textit{ugriz} template light curves.
Another solution for the \textit{g}-band LC peaking before the discovery is also plotted.
{The LCs of SNe 2005gj \citep{aldering06,holtzman08}, 2005kj and 2006aa \citep{taddia13} in \textit{g}-band absolute magnitude are shown in blue, with their phases shifted to have the peak (SN 2005gj) or discovery epoch (SNe 2005kj, 2006aa) matching the SN Ic LC peak.}
{\textit{(Bottom)} Color curves of SNe 2017dio, 2005gj, 2005kj, and 2006aa, in $(g-r)_0$ (corrected for foreground extinction). The green dashed area represents typical SNe Ic \citep{taddia15}.
$(J-H)$ and $(H-K)$ colors are plotted in comparison to SN 2005kj.
}
}
\label{lc}
\end{figure*}

The early light curve (LC) of SN~2017dio was found to be rising (Figure \ref{lc}).
While the presence of narrow H emission features would put SN~2017dio into the SN~IIn subclass, the underlying continuum does not show the typical blue featureless early spectrum \citep[see e.g.][]{kiewe12,taddia13}. At later phases, it became strikingly similar to some SNe IIn (Figure \ref{specs}).
The original SN Ic classification was obtained by omitting these emission lines and only inspecting the underlying spectrum. In the case of SNe IIn, the narrow emission lines are considered as a sign of CSM, whereas the underlying SN in SNe IIn is hidden. SN~2017dio becomes quite clearly a SN IIn after +18 days, but at early phases shows broad absorption/emission features typical of a SE SN. As such, it provides a rare opportunity to investigate the {underlying} SN type {in} SNe IIn. 

\begin{figure*}
\centering
\plottwo{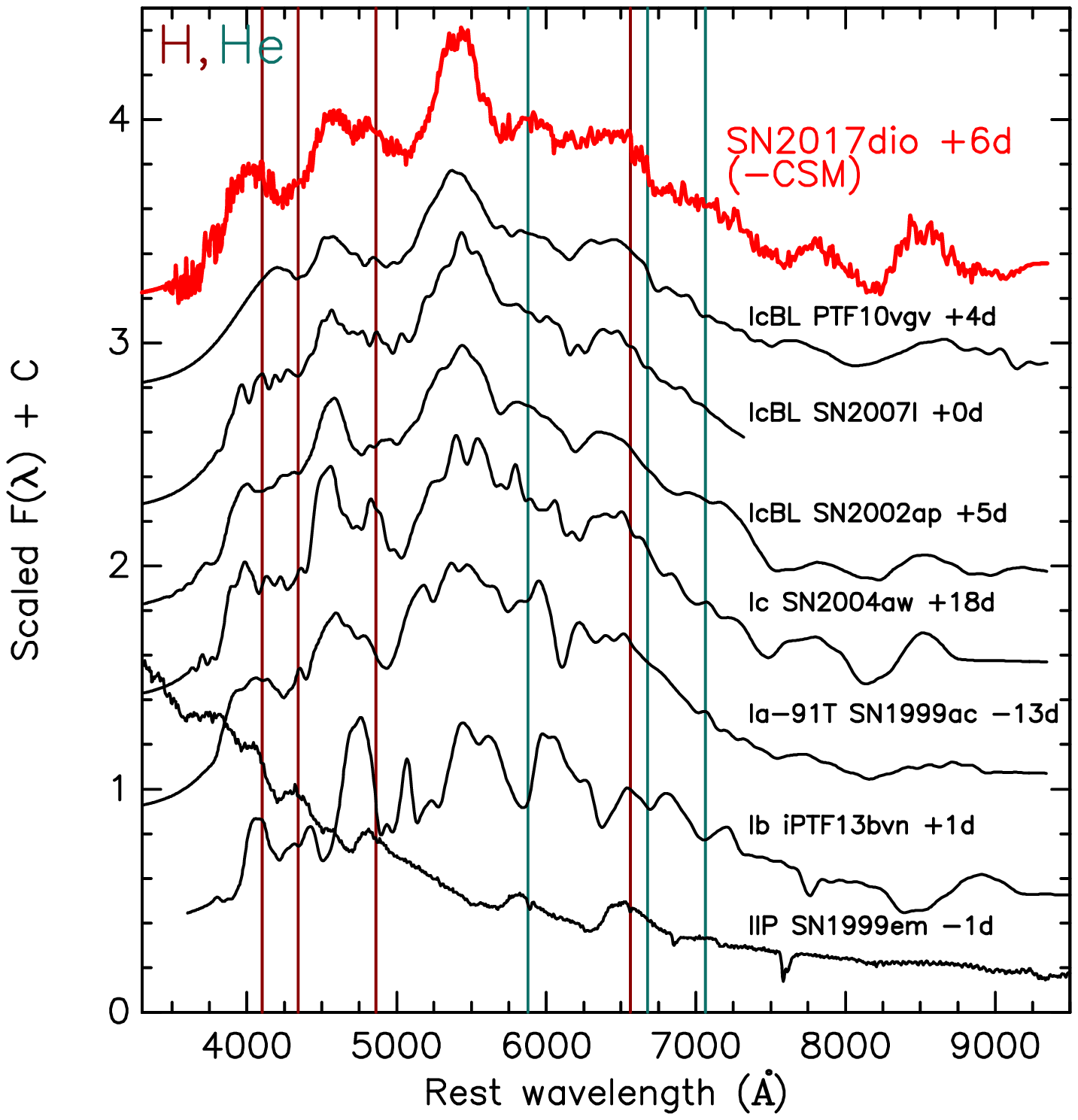}{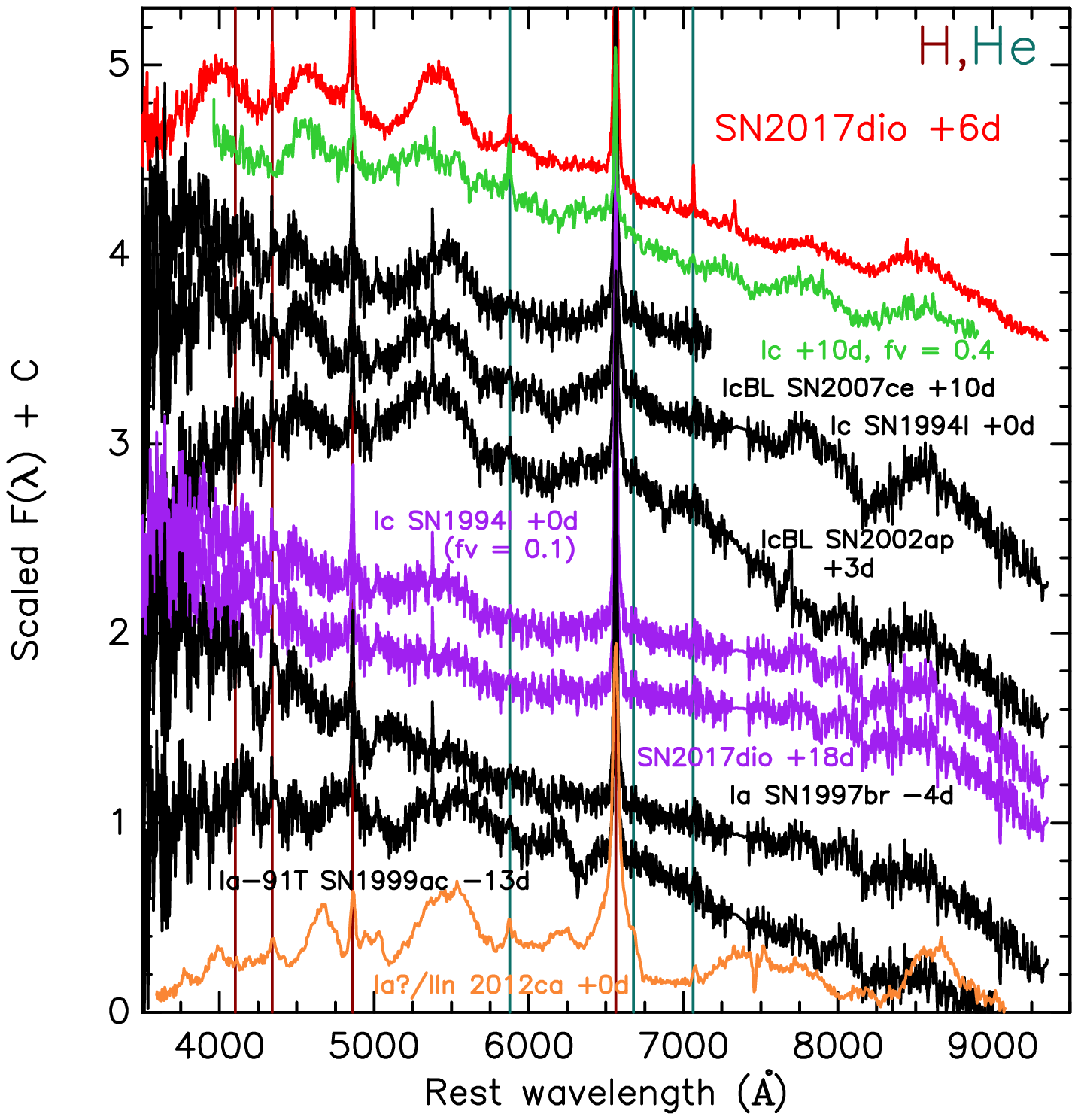}
\caption{
\textit{(Left)} 
{Comparison of {the} CSM interaction-subtracted spectrum of SN~2017dio (red) with {spectra of other} SNe (black; including SNID templates of \citealt{modjaz14}).}
%Normalized mean spectra of SNe Ic and IcBL \citep{liu16,modjaz16} are plotted for comparison. 
\textit{(Right)} 
{Comparison of the +6 days spectrum of SN~2017dio (red) with those of {other} SNe combined with interaction spectrum (black; $f_V = 0.3$), a template SN Ic+CSM interaction spectrum from \citet[][green]{leloudas15}, and SN 2012ca (orange).}
The spectrum of SN 1994I combined with interaction spectrum ($f_V = 0.1$) is showed for comparison with SN 2017dio at +18~days.
The comparison spectra are from WISeREP.
%Normalized mean spectra of SNe Ic and IcBL combined with interaction spectrum are plotted for comparison. 
{In both panels, phases are days from maximum for spectra other than SN~2017dio.}
}
\label{specsub}
\end{figure*}

{
%Alternatively, SN~2017dio could have been a SN Ia interacting with a CSM \citep{silverman13}. 
A number of objects in the literature have been considered as either SN Ia or Ic, such as SNe~2002ic \citep{hamuy03,benetti06} and 2012ca \citep{fox15,inserra16}, due to the spectral similarities of SNe Ia and Ic particularly in the early phase.
Nevertheless, the majority of these objects are believed to be SNe Ia embedded in a dense CSM \citep{silverman13,leloudas15}.
{PTF11kx \citep{dilday12} is considered as a clear example of a bona fide SN Ia with CSM interaction, as the SN was relatively interaction-free in the early phase.}
In Figure \ref{specs}, it can be seen that there are similarities between SN 2017dio at early phase with both SNe Ia and Ic, however no exact match was found and its spectral {appearance} cannot be reproduced by any of those interacting SNe Ia.
}

One way to reassess the classification is by removing the spectral component attributed to the CSM interaction, and then cross-matching the spectrum with a template of SNe with various types. For this purpose, the +18 days spectrum was used as a proxy for the CSM interaction spectrum, and subtracted from the +6 days spectrum to obtain a `pure' SN component. 
Before subtraction, the spectra were normalized by the average flux across the wavelength range.
The first spectrum in Figure \ref{specsub} \textit{(left)} shows the subtraction result, {compared to several other SNe}.
The SNID \citep{blondin07} and GELATO \citep{harutyunyan08} tools were used to compare this spectrum with those of known SNe. Both tools yield a Ic classification around the maximum light, and {a} better match is obtained with broad-lined SN Ic {(hereafter SN IcBL)} spectra {than with} normal SN Ic. 
%This procedure suggested that the CSM-subtracted spectrum of SN~2017dio resembles those of SNe Ic around maximum light.
However, a precise classification is difficult at these epochs \citep{prentice17}.
{Other SN types (Ia, Ib, II) do not provide a better match compared to SN Ic.}

%
%\begin{figure}
%\centering
%\includegraphics[width=\linewidth]{specadd2.eps}
%\caption{Comparison of the +6 day spectrum of SN~2017dio (red) with known SNe Ic combined with CSM interaction spectrum (black), and 
%The SNe Ic spectra are from WISeREP \citep{yaron12}.
%Normalized mean spectra of SNe Ic and IcBL \citep{liu16,modjaz16} combined with interaction spectrum are plotted with khaki and blue. Phases are days from maximum for spectra other than SN~2017dio.}
%\label{specadd}
%\end{figure}

To check the validity of this method, a reverse procedure was applied: the +18 days spectrum representing CSM interaction was coadded to templates of known SN spectra. 
\citet{leloudas15} simulated various pure SN spectra coadded to a CSM interaction component to investigate the spectral identification of SNe IIn, and showed that the flux ratio of the underlying SN to the continuum $f_V$ is the most important parameter determining if an underlying, intrinsic SN spectrum can be classified correctly.
Following \citet{leloudas15}, we experimented with varying $f_V$ to match the appearance of the coadded spectra with the +6 days spectrum.  
We found that the +6 days SN~2017dio spectrum is best matched by type-Ic SNe coadded to the {CSM interaction spectrum with $f_V \gtrsim 0.3$} (Figure \ref{specsub}, \textit{right}).
This is consistent with the finding of \citet{leloudas15} that a critical $f_V \gtrsim 0.3$ is needed in order to correctly classify an underlying pure SN Ic spectrum contaminated by CSM interaction. Below $f_V \approx 0.3$, these events would be indistinguishable from typical SN IIn without a hint of the nature of the underlying SN.

\begin{figure}
\centering
\includegraphics[width=\linewidth]{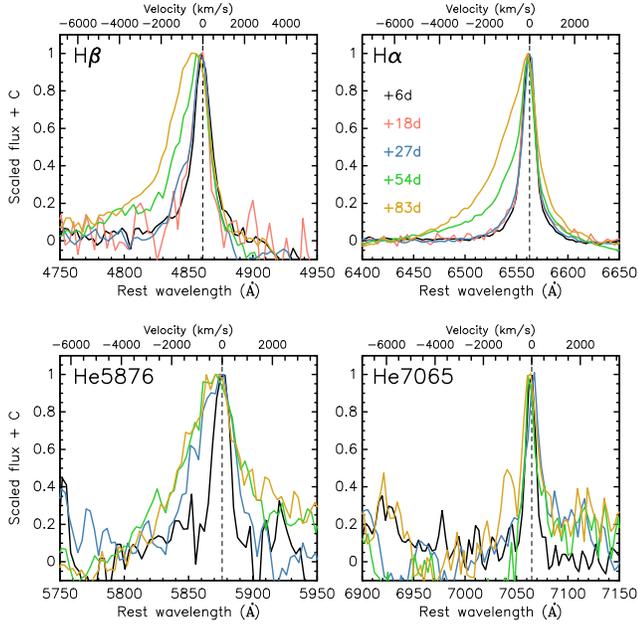}
\caption{The observed line profile evolution for H$\alpha$, H$\beta$, He I $\lambda$5876, He I $\lambda7065$.
Only data taken with the same instrument and setting (ALFOSC) are plotted, for clarity.
}
\label{lineprof}
\end{figure}

\begin{figure}
\centering
\includegraphics[width=\linewidth]{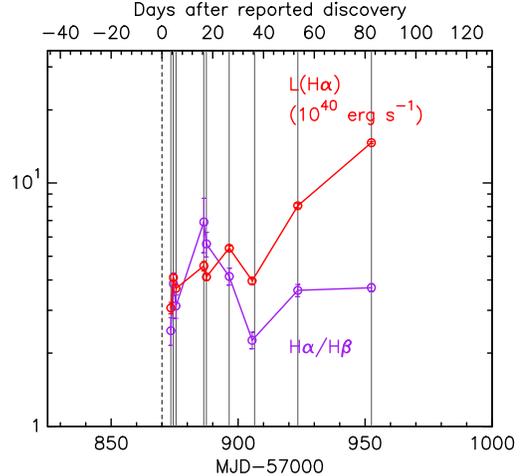}
\caption{The evolution of the H$\alpha$/H$\beta$ flux ratio (purple), and H$\alpha$ line luminosity (red). }
\label{linerat}
\end{figure}

These analyses suggest that SN~2017dio underwent a transition from type Ic to IIn. The SN Ic spectrum was prominent in the early phases, while at later times the CSM interaction component dominates. 
A signature of the increasing strength of CSM interaction is the broadening of the emission lines. Figure~\ref{lineprof} shows the evolution of the strongest H and He lines, showing {their} widths increas{ing} with time. {The} Lorentzian FWHM of the lines was typically around 500 km~s$^{-1}$ at the beginning and increased to $\sim2000$ km~s$^{-1}$ in the later spectra. 
The line profile was initially symmetric with wings indicative of electron scattering \citep[e.g.][]{dessart16}, then evolved into an asymmetric profile with a broad blue component.
{This evolution is similar to that seen in SN 2010jl, where such asymmetric profile may be caused by dust \citep[][]{gall14}, or alternatively radiative acceleration or radiative transfer effects in the cool dense shell \citep{fransson14}. }
%Nevertheless, in the early phase of SN 2017dio such a profile is not observed and the blue color does not support the presence of dust (Figure \ref{lc}). Therefore it could not have affected the early light curve whence the underlying SN Ic peak luminosity is derived.
Nevertheless, in the early phase of SN 2017dio such a profile is not observed.
The observed optical color evolution showing SN 2017dio becoming bluer with time and the H$\alpha$/H$\beta$ line ratio staying relatively constant in these epochs do not support dust formation (Figures \ref{lc} and \ref{linerat}). Furthermore, the observed near-infrared colors are consistent with those expected for a SN IIn without any significant excess emission from dust \citep{taddia13}. 
Therefore, we do not expect dust to affect our observations.
% and consider the interpretation of \citet{fransson14} for the evolution of the line profiles in SN 2010jl to be the likely explanation also in the case of SN 2017dio.

As the spectral evolution points to SN~2017dio being a SN Ic interacting with CSM, the LC should also match this interpretation. The underlying SN Ic luminosity must be fainter compared to the observed LC that includes the CSM interaction contribution. 
In Figure~\ref{lc} we plot template SN Ic LCs from \citet{taddia15} to match our photometry. The epochs of the template LCs are the same, and they are only shifted vertically. It is apparent that the template LC peak corresponds to the epoch of the first spectra. This is consistent with the spectral match to SNe Ic around the peak. In a few days, the observed LC continued rising and there was a flux offset between the observed and template LCs. 
Since then, the subsequent photometric evolution can be explained by the increasing contribution from the interaction. The H$\alpha$ line luminosity is generally increasing, suggesting an increasing strength of the interaction (Figure \ref{linerat}).
Interaction appears to {have} start{ed} about the time of the SN Ic LC peak, yielding extra flux in the observed LC at subsequent phases, and eventually broadening the emission lines.

The observed LC of SN~2017dio peaked at around early July 2017 {(MJD $\sim$ 57930)}, about 60~days after the discovery {and during the interaction dominated phase}. 
{The peak absolute magnitude was $M_g = M_r = -18.8$~mag}. 
\citet{leloudas15} estimated the peak luminosity needed for CSM interaction to hide a SN Ic/IcBL. A SN~IIn needs to {reach} at least $\sim -19$~mag to hide a SN~Ic peaking at $\sim -17.5$~mag.
{In Figure \ref{lc}, the underlying photospheric SN~Ic LC might have reached $M_g = -17.6$~mag at peak, which is consistent with the above scenario.}
A second solution to the LC peak can be placed between the last non-detection and the first photometric point (Figure~\ref{lc}), nevertheless this would yield a similar, perhaps even fainter, peak magnitude.
{The corresponding {photospheric} peak magnitude of $M_r = -18.1$~mag is well within the observed distributions of SNe Ic and IcBL \citep[$M_R = -18.5 \pm 0.8$ mag and $M_R = -19.0 \pm 1.1$ mag, respectively;][]{drout11}.}
SNe Ia with CSM interaction are  found to peak brighter than $M_V \sim -19.5$ mag \citep{leloudas15,silverman13}, which is constrained by the underlying SN Ia LC peaking between $-18.5$ and $-19.7$ mag. These luminosity ranges cannot accommodate both the observed LC peak and the underlying SN Ic {photospheric} LC peak in SN~2017dio. 
{In SNe Ia with CSM interaction, the underlying SN is always spectroscopically similar to the SN 1991T-like objects, which is a brighter subgroup of SNe Ia.
They are spectroscopically distinguishable from SN Ic \citep[e.g.][]{leloudas15} and the spectra of SN 2017dio do not show similarities with these objects.}
The extinction towards SN~2017dio appears to be negligible (i.e. no sign of Na~I absorption or high Balmer decrement, and normal colors), ruling out a higher intrinsic luminosity that would be required by the scenario for a SN Ia with CSM interaction.

The consistency between the spectral and photometric evolutions suggests that the interpretation of SN~2017dio being a SN Ic interacting with a CSM and transitioning into a SN IIn is robust.
%\footnote{Several SNe e.g. 2002ic and 2012ca were argued to be either SNe~Ia \citep{hamuy03,fox15} or Ic \citep{benetti06,inserra16} interacting with CSM.}. 
{Alternative explanations such as SN~Ia, II, or Ib with CSM interaction are not plausible.}

%interacting with CSM would require a significantly higher luminosity \citep[$M_V < -19.5$ mag,][]{leloudas15}, and there would be negligible host galaxy reddening to account for the offset (e.g. no sign of Na~I absorption or high Balmer decrement).
%The observed spectrum also cannot be explained with a SN~II or Ib. 

%Finally, it is intriguing to investigate whether many SNe IIn are actually similar to SN~2017dio, and we are simply missing their early evolution. This is possible, since most SNe IIn have been discovered close to or after the maximum light with the CSM interaction being already dominant. 

\subsection{Host galaxy}

The field of SN~2017dio is in the SDSS database. The host galaxy, SDSS J113627.76+181747.3, has $u=23.36$, $g=21.12$, $r=20.67$, $i=20.55$, and $z=20.12$ mag. Considering the distance modulus and foreground reddening, the absolute magnitude is $M_g = -15.1$ mag. The galaxy appears to be small without a discernible shape. Its diameter of $\sim3$" corresponds to $\sim2$ kpc at the distance of $\sim160$ Mpc. This suggests that the host is a dwarf galaxy smaller and fainter than the Small Magellanic Cloud \citep[$M_V = -16.8$ mag, ][]{mcconachie12}.

In the spectra of SN~2017dio, the interstellar emission lines from the galaxy are hidden by the SN. Therefore, no strong-line analyses such as estimates of star formation rate or metallicity can be done. Employing the galaxy luminosity-metallicity relation of \citet{tremonti04} yields an estimate of 12+log(O/H) $\approx8.0$ dex for the galaxy.
Comparing this luminosity to the sample of SN Ic and IcBL hosts \citep{modjaz08}, the host of SN~2017dio falls at the faint end of the distribution. Dwarf galaxies with the size of SN~2017dio host predominantly produce SNe~IcBL rather than normal SNe~Ic \citep{arcavi10}. This is {also consistent} with the possibility that SN~2017dio is a SN~IcBL.

\subsection{CSM and progenitor characteristics}

The spectra and LC show that the CSM interaction was relatively weak at the earliest epochs.
Therefore, the bulk of the CSM is not located {at immediate vicinity of} the progenitor star. The SN ejecta must have traveled outward in a rarefied environment before encountering the denser CSM parts. The H$\alpha$/H$\beta$ line ratio was around 3 in the early phases and {two times} higher after two weeks, consistent with increasing CSM densities.
After this peak, the H$\alpha$/H$\beta$ ratio drops and rises again indicating fluctuations in the CSM density (Figure \ref{linerat}). 
In a $\rho \sim r^{-2}$ spherically-distributed CSM created by steady-state stellar winds \citep{cf94}, such behavior is not expected. 
%However, the mechanisms responsible for the observed ratio may be different in the earlier and later epochs.
% thus the observed values between the pre-interaction and interaction-dominated phases are not directly comparable.

%The spectral evolution may be explained as follows. 
At the early epochs, the emission lines are narrow with symmetric wings. They are likely to arise from continuous ionization by the interaction of the shock with the CSM, and {effects of electron scattering within the CSM}. 
The later spectra show asymmetric lines with broad blue wings (Figure \ref{lineprof}). 
%CSM interaction is the dominant ionization source at this stage.
The asymmetric line profile may be caused by the interaction region being occulted at the {receding} side of the optically thick {ejecta}, thus causing the red wing to be suppressed. 
At the earliest epochs the CSM interaction was weak, and thus the SN~Ic spectral characteristics were visible. Later, when the CSM interaction dominates, the SN~Ic features are hidden by the CSM-interaction component.

Thus, the geometry of the CSM embedding SN~2017dio progenitor does not seem to be consistent with that generated from {spherically symmetric} mass loss via winds. 
Instead, it could have a clumpy, {toroidal}, or even bipolar distribution.
%In either case, an increase of CSM density with distance from the progenitor is required to explain the SN evolution. 
With the averaged {expansion} velocity of 10000 km~s$^{-1}$ {for SNe Ib/c} \citep[e.g.][]{cano13}, the SN ejecta took 80 days to reach the part of the CSM corresponding to the peak H$\alpha$ luminosity. This translates into a distance of $\sim500$~AU. 
Presumably, this part of the CSM was created through mass loss with velocity of 500 km~s$^{-1}$.
The luminosity of the H$\alpha$ line can be used to estimate the mass loss rate of the progenitor star responsible for the CSM build-up in the vicinity, as H$\alpha$ luminosity is proportional to the kinetic energy dissipated per unit time.
{The observed peak H$\alpha$ luminosity of SN~2017dio is $\sim 1.5 \times 10^{41}$ erg~s$^{-1}$.}
{The peak H$\alpha$ luminosity can be used to estimate the progenitor mass loss rate by employing the relation}
\begin{equation}
\dot{M} = 2 {{L_{H\alpha}}\over{\epsilon_{H\alpha}}} {{v_\textrm{wind}}\over{(v_\textrm{shock})^3}}
\end{equation}

{Assuming a progenitor wind velocity of 500 km~s$^{-1}$ from emission line FWHM at the early epochs, and an efficiency factor $\epsilon_{H\alpha}$ of 0.01 \citep{cf94}, the pre-SN mass loss rate of the progenitor star was estimated to be $\sim 0.02$ $(\epsilon_{H\alpha}/0.01)^{-1}$ $(v_\textrm{wind}/500$ km s$^{-1}$) $(v_\textrm{shock}/10 000$ km s$^{-1})^{-3}$ $M_\odot$~yr$^{-1}$. }
The shock velocity must be lower compared to the velocity of the part of the ejecta that is not interacting with the CSM, but this is hidden by the electron scattering wings which extend up to $\sim 10000$ km~s$^{-1}$ (Figure \ref{lineprof}). 
{The derived progenitor mass loss rate is comparable to some SNe~IIn \citep{kiewe12,taddia13}.
It is a few times lower compared to SN 2010jl ($\dot{M} \sim 0.1$ $M_\odot$~yr$^{-1}$, and consistently, so are the peak bolometric\footnote{{Calculated from the $g$-band light curve and $(g-r)$ colors using the simple prescription of \citet{lyman14}.}} 
and H$\alpha$ luminosities ($\sim10^{43}$ erg s$^{-1}$ and $\sim 10^{41}$ erg s$^{-1}$, respectively, compared to $\sim 3\times10^{43}$ erg~s$^{-1}$ and $\sim 10^{42}$ erg~s$^{-1}$; \citealt{fransson14}).
}

The progenitor must {therefore} have experienced major mass loss a few decades before the SN. {Then,} the rest of the envelope was removed through less vigorous mass loss in the final years, until being removed completely prior to the SN. This is reminiscent to the inferred progenitor behavior of the type-Ib/IIn SN~2014C \citep{milisav15}.
High pre-SN mass loss may be caused by eruptions \citep[e.g.][]{smith14} or stripping by a close binary companion \citep[e.g.][]{yoon17}.
%In the case of binary stripping, following mass transfer the mass loss rate may be increased until the complete removal of H, and the subsequent removal of He layer is powered by winds at a somewhat lower mass loss rate \citep{yoon17}.
With a wind-driven mass loss, it is inconceivable for the progenitor to still retain the H-rich material nearby while the He layer is depleted, as observed in SN 2017dio.
In the binary progenitor scenario, the primary transfers material to the secondary through a Roche-lobe overflow and evolves into a C+O star. With an increased mass, the secondary's evolution is accelerated and it leaves the main sequence before the explosion of the primary. The secondary then experiences a luminous blue variable (LBV) phase, or becomes a giant and starts a reverse mass transfer to the primary. At this point there is a large amount of H-rich CSM around the primary and it explodes as a SN Ic. 
This path could have occurred within the standard binary evolution, and the progenitor system may be somewhat similar to some Wolf-Rayet+LBV binaries such as HD~5980 \citep{koenig14}.
%and perhaps is not very different from 1993J \citep{nomoto93}.
%In comparison with SNe Ibn, the prototypical SN 2006jc experienced a giant outburst two years prior to the SN \citep{pastorello07}, and a possible lower-mass companion star has been detected \citep{maund16}.
{The prototypical type Ibn SN 2006jc experienced a giant outburst two years prior to the SN \citep{pastorello07}. A possible lower-mass companion star detected in late-time observations together with the lack of further outbursts after the explosion suggest that the outburst originated from the SN progenitor itself \citep{maund16}.}
Given that SN 2017dio could be a SN IcBL, we note that binary evolution may also play an important role for SNe IcBL.

\section{Summary}

The early spectrum of SN~2017dio resembles those of SNe Ic, with a CSM component.
%Later, the CSM interaction component grew stronger and gradually hid the SN Ic spectrum. 
%Subsequent spectral evolution resembles that of typical SNe IIn. 
SN~2017dio appears to be brightening after {the} discovery, accompanied by an increase of H$\alpha$ luminosity and a broadening of the emission lines. This is consistent with CSM interaction becoming dominant.
The evolution suggests that the epoch of maximum for the underlying SN Ic was around {the} discovery. With this constraint, the peak luminosity of SN~2017dio falls into the range commonly observed for SNe Ic.

The CSM must not have been distributed as $\rho \sim r^{-2}$, typical for a steady-state stellar wind. This would require that the progenitor star underwent vigorous mass-loss episodes a few decades before the SN, possibly driven by eruption or binary interaction.

%Finally, we encourage efforts from the community to observe SN~2017dio in all accessible wavelengths, to monitor its evolution as the ejecta sweeps through the CSM, allowing probing of previous mass loss episodes.

\acknowledgments

{We thank the referee and Sung-Chul Yoon for useful suggestions}.
KM acknowledges FINCA visitor program, Japan Society for the Promotion of Science (JSPS) through KAKENHI Grant 17H02864 and JSPS Open Partnership Bilateral Joint Research Project between Japan and Chile.
LG was supported by the US National Science Foundation, Grant AST-1311862.
CPG acknowledges EU/FP7-ERC grant No. [615929].
AP, SB, and NER are supported by the PRIN-INAF 2014 project Transient Universe: unveiling new types of stellar explosions with PESSTO.
Support for GP is provided by the Ministry of Economy, Development, and Tourism's Millennium Science Initiative through grant IC120009, awarded to The Millennium Institute of Astrophysics, MAS.
CG is supported by the Carlsberg Foundation. 
MDS is supported by a research grant (13261) from the VILLUM FONDEN. NUTS is supported in part by IDA (Instrumentcenter for Danish Astrophysics).
The ATLAS surveys are funded through NASA grants NNX12AR55G.
 

\vspace{5mm}
%\facilities{HST(STIS), Swift(XRT and UVOT), AAVSO, CTIO:1.3m, CTIO:1.5m,CXO}
\facilities{NOT(ALFOSC,NOTCam), NTT(EFOSC2), LT(SPRAT), LCOGT: 2m}


\begin{thebibliography}{}

%\bibitem[Astropy Collaboration et al.(2013)]{2013A&A...558A..33A} Astropy Collaboration, Robitaille, T.~P., Tollerud, E.~J., et al.\ 2013, \aap, 558, A33 
%\bibitem[Bertin \& Arnouts(1996)]{1996A&AS..117..393B} Bertin, E., \& Arnouts, S.\ 1996, \aaps, 117, 393 
%\bibitem[Corrales(2015)]{2015ApJ...805...23C} Corrales, L.\ 2015, \apj, 805, 23
%\bibitem[Ferland et al.(2013)]{2013RMxAA..49..137F} Ferland, G.~J., Porter, R.~L., van Hoof, P.~A.~M., et al.\ 2013, \rmxaa, 49, 137
%\bibitem[Hanisch \& Biemesderfer(1989)]{1989BAAS...21..780H} Hanisch, R.~J., \& Biemesderfer, C.~D.\ 1989, \baas, 21, 780 
%\bibitem[Lamport(1994)]{lamport94} Lamport, L. 1994, LaTeX: A Document Preparation System, 2nd Edition (Boston, Addison-Wesley Professional)
%\bibitem[Schwarz et al.(2011)]{2011ApJS..197...31S} Schwarz, G.~J., Ness, J.-U., Osborne, J.~P., et al.\ 2011, \apjs, 197, 31  
%\bibitem[Vogt et al.(2014)]{2014ApJ...793..127V} Vogt, F.~P.~A., Dopita, M.~A., Kewley, L.~J., et al.\ 2014, \apj, 793, 127  

%\bibitem[Abolfathi et al.(2017)]{abolfathi17} Abolfathi, B., Aguado, D.~S., Aguilar, G., et al.\ 2017, arXiv:1707.09322 
%\bibitem[Anderson et al.(2012)]{anderson12} Anderson, J.~P., Habergham, S.~M., James, P.~A., \& Hamuy, M.\ 2012, \mnras, 424, 1372 
\bibitem[Aldering et al.(2006)]{aldering06} Aldering, G., Antilogus, P., Bailey, S., et al.\ 2006, \apj, 650, 510 
\bibitem[Arcavi et al.(2010)]{arcavi10} Arcavi, I., Gal-Yam, A., Kasliwal, M.~M., et al.\ 2010, \apj, 721, 777 
\bibitem[Ben-Ami et al.(2014)]{benami14} Ben-Ami, S., Gal-Yam, A., Mazzali, P.~A., et al.\ 2014, \apj, 785, 37 
\bibitem[Benetti et al.(2006)]{benetti06} Benetti, S., Cappellaro, E., Turatto, M., et al.\ 2006, \apjl, 653, L129 
\bibitem[Blondin \& Tonry(2007)]{blondin07} Blondin, S., \& Tonry, J.~L.\ 2007, \apj, 666, 1024 
\bibitem[Buzzoni et al.(1984)]{buzzoni84} Buzzoni, B., Delabre, B., Dekker, H., et al.\ 1984, The Messenger, 38, 9 
\bibitem[Cano(2013)]{cano13} Cano, Z.\ 2013, \mnras, 434, 1098 
\bibitem[Cartier et al.(2017)]{cartier17} Cartier, R., Gutierrez, C.~P., Smith, M., et al.\ 2017, The Astronomer's Telegram, 10334
\bibitem[Chevalier \& Fransson(1994)]{cf94} Chevalier, R.~A., \& Fransson, C.\ 1994, \apj, 420, 268 
\bibitem[Chugai \& Chevalier(2006)]{chugai06} Chugai, N.~N., \& Chevalier, R.~A.\ 2006, \apj, 641, 1051 
\bibitem[Dessart et al.(2016)]{dessart16} Dessart, L., Hillier, D.~J., Audit, E., Livne, E., \& Waldman, R.\ 2016, \mnras, 458, 2094 
\bibitem[Dilday et al.(2012)]{dilday12} Dilday, B., Howell, D.~A., Cenko, S.~B., et al.\ 2012, Science, 337, 942 
\bibitem[Drake et al.(2009)]{drake09} Drake, A.~J., Djorgovski, S.~G., Mahabal, A., et al.\ 2009, \apj, 696, 870 
\bibitem[Drout et al.(2011)]{drout11} Drout, M.~R., Soderberg, A.~M., Gal-Yam, A., et al.\ 2011, \apj, 741, 97   
%\bibitem[Filippenko(1997)]{filippenko97} Filippenko, A.~V.\ 1997, \araa, 35, 309 
\bibitem[Fox et al.(2015)]{fox15} Fox, O.~D., Silverman, J.~M., Filippenko, A.~V., et al.\ 2015, \mnras, 447, 772 
\bibitem[Fransson et al.(2014)]{fransson14} Fransson, C., Ergon, M., Challis, P.~J., et al.\ 2014, \apj, 797, 118 
%\bibitem[Galbany et al.(2016)]{galbany16} Galbany, L., Stanishev, V., Mour{\~a}o, A.~M., et al.\ 2016, \aap, 591, A48 
\bibitem[Gall et al.(2014)]{gall14} Gall, C., Hjorth, J., Watson, D., et al.\ 2014, \nat, 511, 326 
\bibitem[Gal-Yam(2016)]{galyam16} Gal-Yam, A.\ 2016, in Handbook of Supernovae, eds. Alsabti, A. W. \& Murdin, P., Springer (arXiv:1611.09353)
\bibitem[Hamuy et al.(2003)]{hamuy03} Hamuy, M., Phillips, M.~M., Suntzeff, N.~B., et al.\ 2003, \nat, 424, 651 
\bibitem[Guillochon et al.(2017)]{guillochon17} Guillochon, J., Parrent, J., Kelley, L.~Z., \& Margutti, R.\ 2017, \apj, 835, 64 
\bibitem[Harutyunyan et al.(2008)]{harutyunyan08} Harutyunyan, A.~H., Pfahler, P., Pastorello, A., et al.\ 2008, \aap, 488, 383 
\bibitem[Holtzman et al.(2008)]{holtzman08} Holtzman, J.~A., Marriner, J., Kessler, R., et al.\ 2008, \aj, 136, 2306 
\bibitem[Inserra et al.(2016)]{inserra16} Inserra, C., Fraser, M., Smartt, S.~J., et al.\ 2016, \mnras, 459, 2721 
%\bibitem[Kangas et al.(2017)]{kangas17} Kangas, T., Portinari, L., Mattila, S., et al.\ 2017, \aap, 597, A92 
\bibitem[Kiewe et al.(2012)]{kiewe12} Kiewe, M., Gal-Yam, A., Arcavi, I., et al.\ 2012, \apj, 744, 10 
\bibitem[Koenigsberger et al.(2014)]{koenig14} Koenigsberger, G., Morrell, N., Hillier, D.~J., et al.\ 2014, \aj, 148, 62 
%\bibitem[Kuncarayakti et al.(2013)]{hk13a} Kuncarayakti, H., Doi, M., Aldering, G., et al.\ 2013a, \aj, 146, 30 
\bibitem[Leloudas et al.(2015)]{leloudas15} Leloudas, G., Hsiao, E.~Y., Johansson, J., et al.\ 2015, \aap, 574, A61 
%\bibitem[Liu et al.(2016)]{liu16} Liu, Y.-Q., Modjaz, M., Bianco, F.~B., \& Graur, O.\ 2016, \apj, 827, 90 
\bibitem[Lyman et al.(2014)]{lyman14} Lyman, J.~D., Bersier, D., \& James, P.~A.\ 2014, \mnras, 437, 3848 
\bibitem[Maund et al.(2016)]{maund16} Maund, J.~R., Pastorello, A., Mattila, S., Itagaki, K., \& Boles, T.\ 2016, \apj, 833, 128 
\bibitem[Mattila et al.(2016)]{mattila16} Mattila, S., Elias-Rosa, N., Lundqvist, P., et al.\ 2016, The Astronomer's Telegram, 8992
\bibitem[McConnachie(2012)]{mcconachie12} McConnachie, A.~W.\ 2012, \aj, 144, 4 
\bibitem[Milisavljevic et al.(2015)]{milisav15} Milisavljevic, D., Margutti, R., Kamble, A., et al.\ 2015, \apj, 815, 120 
\bibitem[Modjaz et al.(2008)]{modjaz08} Modjaz, M., Kewley, L., Kirshner, R.~P., et al.\ 2008, \aj, 135, 1136 
\bibitem[Modjaz et al.(2014)]{modjaz14} Modjaz, M., Blondin, S., Kirshner, R.~P., et al.\ 2014, \aj, 147, 99  
%\bibitem[Modjaz et al.(2016)]{modjaz16} Modjaz, M., Liu, Y.~Q., Bianco, F.~B., \& Graur, O.\ 2016, \apj, 832, 108 
%\bibitem[Nomoto et al.(1993)]{nomoto93} Nomoto, K., Suzuki, T., Shigeyama, T., et al.\ 1993, \nat, 364, 507 
%\bibitem[Patat et al.(1995)]{patat95} Patat, F., Chugai, N., \& Mazzali, P.~A.\ 1995, \aap, 299, 715 
\bibitem[Pastorello et al.(2007)]{pastorello07} Pastorello, A., Smartt, S.~J., Mattila, S., et al.\ 2007, \nat, 447, 829 
\bibitem[Piascik et al.(2014)]{piascik14} Piascik, A.~S., Steele, I.~A., Bates, S.~D., et al.\ 2014, \procspie, 9147, 91478H 
%\bibitem[Planck Collaboration(2016)]{planck16} Planck Collaboration, Ade, P.~A.~R., Aghanim, N., et al.\ 2016, \aap, 594, A13 
\bibitem[Podsiadlowski et al.(1992)]{podsiadlowski92} Podsiadlowski, P., Joss, P.~C., \& Hsu, J.~J.~L.\ 1992, \apj, 391, 246 
\bibitem[Prentice \& Mazzali(2017)]{prentice17} Prentice, S.~J., \& Mazzali, P.~A.\ 2017, \mnras, 469, 2672 
\bibitem[Roy et al.(2016)]{roy16} Roy, R., Sollerman, J., Silverman, J.~M., et al.\ 2016, \aap, 596, A67 
%\bibitem[Salamanca et al.(1998)]{salamanca98} Salamanca, I., Cid-Fernandes, R., Tenorio-Tagle, G., et al.\ 1998, \mnras, 300, L17 
\bibitem[Schlafly \& Finkbeiner(2011)]{schlafly11} Schlafly, E.~F., \& Finkbeiner, D.~P.\ 2011, \apj, 737, 103 
\bibitem[Schlegel(1990)]{schlegel90} Schlegel, E.~M.\ 1990, \mnras, 244, 269 
\bibitem[Silverman et al.(2013)]{silverman13} Silverman, J.~M., Nugent, P.~E., Gal-Yam, A., et al.\ 2013, \apjs, 207, 3 
\bibitem[Smartt et al.(2015)]{smartt15} Smartt, S.~J., Valenti, S., Fraser, M., et al.\ 2015, \aap, 579, A40 
\bibitem[Smith \& Arnett(2014)]{smith14} Smith, N., \& Arnett, W.~D.\ 2014, \apj, 785, 82 
\bibitem[Taddia et al.(2013)]{taddia13} Taddia, F., Stritzinger, M.~D., Sollerman, J., et al.\ 2013, \aap, 555, A10 
\bibitem[Taddia et al.(2015)]{taddia15} Taddia, F., Sollerman, J., Leloudas, G., et al.\ 2015, \aap, 574, A60 
\bibitem[Tonry(2011)]{tonry11} Tonry, J.~L.\ 2011, \pasp, 123, 58 
\bibitem[Tremonti et al.(2004)]{tremonti04} Tremonti, C.~A., Heckman, T.~M., Kauffmann, G., et al.\ 2004, \apj, 613, 898 
%\bibitem[Trundle et al.(2009)]{trundle09} Trundle, C., Pastorello, A., Benetti, S., et al.\ 2009, \aap, 504, 945 
%\bibitem[Turatto et al.(2000)]{turatto00} Turatto, M., Suzuki, T., Mazzali, P.~A., et al.\ 2000, \apjl, 534, L57 
\bibitem[Vink et al.(2001)]{vink01} Vink, J.~S., de Koter, A., \& Lamers, H.~J.~G.~L.~M. 2001, \aap, 369, 574  
\bibitem[Vinko et al.(2017)]{vinko17} Vinko, J., Pooley, D., Silverman, J.~M., et al.\ 2017, \apj, 837, 62 
%\bibitem[Yan et al.(2015)]{yan15} Yan, L., Quimby, R., Ofek, E., et al.\ 2015, \apj, 814, 108 
\bibitem[Yan et al.(2017)]{yan17} Yan, L., Lunnan, R., Perley, D.~A., et al.\ 2017, \apj, 848, 6 
\bibitem[Yaron \& Gal-Yam(2012)]{yaron12} Yaron, O., \& Gal-Yam, A.\ 2012, \pasp, 124, 668 
\bibitem[Yoon et al.(2017)]{yoon17} Yoon, S.-C., Dessart, L., \& Clocchiatti, A.\ 2017, \apj, 840, 10 


\end{thebibliography}
\end{document}